# Stimulated low frequency Raman scattering in high ordered solid nanostructures

*M.V. Tareeva[1], V. A. Dravin[1], R. A. Khmelnitsky [1,2,3,4], K. A. Tsarik[5]*

[1] Lebedev Physical Institute, Russian Academy of Sciences, Moscow, Russia, 119991

[2] Kotel'nikov Institute of Radio Engineering and Electronics, Russian Academy of Sciences, Fryazino Branch, Fryazino, Moscow region, Russia, 141190

[3] Prokhorov General Physics Institute, Russian Academy of Sciences, Moscow, Russia, 119991

[4] Troitsk Institute for Innovation and Fusion Research (GNTs RF TRINITI), Troitsk, Moscow region, Russia, 142190

[5] National Research University of Electronic Technology, Zelenograd, Moscow, Russia, 124498

[*] Corresponding author e-mail:    tareeva@sci.lebedev.ru

**Index topic**

VIII. Nonlinear optics

**Highlights**

- Experimental observation of coherent phonon-mode excitation with high efficiency in submicron nanostructures of a different nature is reported
- Nonlinear mechanism of pumping light scattering on the eigen acoustic vibration of nanostructured medium is proposed.

**Abstract**

Stimulated low frequency Raman scattering (SLFRS) in single-crystal diamond films (SCD) with a graphitized layer built-in is investigated. SLFRS process in all studied structures has frequency shift in gigahertz range (5.1-13.2 GHz). Coherent phonon mode excitation at a frequency of several GHz is a result of nonlinear interaction of high-power laser wave with submicron films. SLFRS conversion efficiency and threshold is estimated experimentally. Comparison of the obtained results with data on SLFRS properties in synthetic opal matrixes is made.

## 1   Introduction

Stimulated low frequency Raman scattering (SLFRS) is a fundamental effect caused by laser pulse interaction with acoustic vibrations of nanostructured medium [1, 2].
Since the first experimental observation of spontaneous low frequency Raman scattering (LFRS) in 1986 [1], it is used as an effective research method of nanosized systems.
The frequency shift of SLFRS is defined by eigen frequencies of scattering centers of the system and, hence, gives important information on morphology of the system under study [2].
SLFRS which is accompanied by coherent excitation of acoustic oscillations can be excited in materials with different structure, composition and morphology [3- 5]. Wide range of inorganic [3-5] and organic substances [6], high-ordered [3] and random [4] materials has been characterized by low frequency Raman scattering in spontaneous and stimulated regime [5].

The main SLFRS features are high conversion efficiency, narrow width of the spectral line and divergence close to the laser one. These features are the manifestation of the coherent excitation of the system at the nanoscale by power laser wave. The nature of such coherent behavior lies in coherent oscillations of the dipole moments, induced by the powerful electromagnetic laser field. Such a mechanism is the same for both the processes of SLFRS and stimulated Raman scattering, which is driven by coherently excited lattice or molecular vibrations [7].

One of the most convenient objects for classical demonstration of nonlinear mechanism of SLFRS are synthetic opals [8-10], which are colloidal crystal composed of regularly packed monodisperse submicron globules of silicon dioxide ($SiO_2$). This artificial material is called synthetic opal due to its opalescence that appears because of periodic dielectric structure with the crystal lattice constant close to the wavelength of the visible light.

Due to the spatial modulation of the optical and acoustic properties (permittivity and Lame constants), synthetic opals possess photonic-band gaps and phonon modes quantization [11].

Modifications of the photonic density of states in some regions lead to the large values of the electromagnetic field localization in synthetic opals, and, hence, to the strong enhancement of nonlinear wave-matter interaction [8]. Residual pores in colloidal structure of synthetic opal could be relatively easily filled with various test substances, for example by organic liquids, which makes synthetic opals convenient for investigation various types of nonlinear light scatterings, primarily Raman scattering [9,11].

In this work we present new data related to low frequency coherent excitation in single-crystal diamond films (SCD) with a graphitized layer built-in. These data are compared with the parameters of SLFRS excited in synthetic opal matrixes.

## 2. Preparation and characterization of the samples

*2.1 Single-Crystal Diamond Films*

The diamond samples were cut from polished plates of natural single crystals (nitrogen impurity content less than $5 \cdot 10^{18}$ cm$^{-3}$) parallel to the {110} crystallographic plane and have been subjected to the further ion implantation and annealing [12].

While ion implantation always results in radiation damage of material, following annealing of heavily damaged diamond leads to the formation of a graphitized layer in the region of maximum radiation influence, where carbon atoms are bounded by $sp^2$ bonds [12, 13].

The polished samples of SCD plates in our experiment have been bombarded at room temperature by $^{12}C^+$, $^4He^+$ and $D^+$ ions with energies of 50-350 keV (see Table 1), than annealed for one hour at temperatures 1480 °C (1600°C for 350 keV $D^+$ sample) in vacuum at $10^{-3}$ Pa in a graphite wall furnace. Surface graphitization was etched out using $H_2SO_4$ + $K_2Cr_2O_7$ solution at a temperature of about 180° C.

Such treatments lead to graphitization of buried submicron layer in a bulk SCD, which is covered by restored submicron SCD [13]. The depth of graphitization, i.e. the thickness of the buried graphitized layer is defined by a set of factors such as the ion implantation energy and dose, as well as the type of bombarding ions [12]. For representative results we used three types of ion and varied implantation regimes (see Table 1). As a result we have SCD films with thin submicron ion-beam-induced graphitized layer of a certain thickness buried on a certain depth in each sample (see Table 1).

Actual parameters of the buried graphitized layer in the bulk material could be controlled by the type and the energy of bombarding ions and by the implantation dose. The position and thickness of graphitized layer have been estimated considering the radiation damage profile of the sample using Monte Carlo technique [14].

Table 1. Ion implantation regimes for pristine samples of natural SCD films and parameters of a buried graphitized layer.

| Sample name | Type of ion | Ion energy, keV | Implantation dose, $10^{16}$ см$^{-2}$ | Restored SCD layer depth, nm | Graphitized layer thickness, nm | SLFRS frequency shift*, $GHz$ | SLFRS frequency shift*, см$^{-1}$ |
|---|---|---|---|---|---|---|---|
| 50 keV $^4$He$^+$ | $^4$He$^+$ | 50 | 2.5 | 185 | 84 | 9.3 | 0.31±0.01 |
| 350 keV $^{12}$C$^+$ | $^{12}$C$^+$ | 350 | 0.4 | 365 | 100 | 9 | 0.30±0.01 |
| 350 keV D$^+$ | D$^+$ | 350 | 12 | 1680 | 150 | 8.4 | 0.28±0.03 |

*30°-60° scattering geometry

The film surface image of SDC film made by atomic force microscopy (AFM) in contact mode of surface scanning is presented in Figure 1.

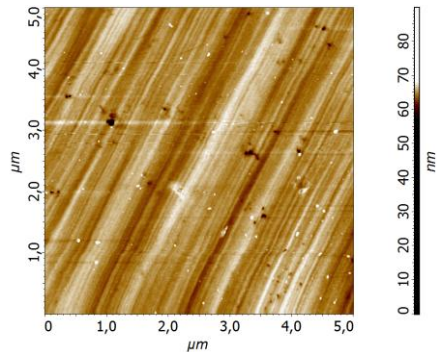

Fig.1. (a) 5x5 μm AFM image of the surface {110} of the submicron SCD films implanted with $^4$He$^+$ ions at the energy of 50 keV to the dose of $2.5\times 10^{16}$ cm$^{-2}$ in two-dimensional view. Horizontal and vertical axes are given in μm, gradient scale are given in nm. Graphitized layer is buried on depth $h$ (see Table.2) under the surface and has boundaries with roughness of about 7 nm with a bulk SCD (300 μm -thickness). The thickness of the buried graphitized layer is ranged from 80 to 100 nm. Surface swelling above the buried layer is ranged from 20 to 100 nm in different samples.

This ATM data show nanoscale wave-type grooves, which remained after mechanical treatment, that evidence about well done mechanical polishing which indicates a good quality of mechanical polishing: roughness of the film surface about 1.18 nm.

*2.2 Synthetic Opal Matrices*

When creating samples of synthetic opals, first a synthesis of a monodisperse suspension of spherical silica particles (SiO$_2$ globules) was carried out by hydrolysis of tetraethoxysilane in ethanol in the presence of ammonia.

The resulting globules can be ordered in various ways; one of the most common practices of the design of synthetic opals is the sedimentation of globules under the influence of gravity (see, for example, [15,16]). Natural sedimentation was used in the present work as a main method, which practically has no limitations on the size of the samples produced. The samples of opals used in this work were about 5x5x1 mm$^3$.

The resulting structures after drying were heat-treated to strengthen the structure by sintering the SiO$_2$ globules. As a result of such treatment SiO$_2$ globule array were strengthen by the formation of strong siloxane bonds (Si-O-Si) between the globules. In the end of all processes we have a solid structure, formed by densely packed layers of identical SiO$_2$ globules with a diameter of several hundred nm with octahedral and tetrahedral pores between them [17].

The characteristics of synthetic opal matrices used in the present experiment are summarized in Table 2.

Table 2. Optical and geometric characteristics of synthetic opal matrices used in the experiment and corresponding SLFRS frequencies.

| Sample name | Diameter of SiO$_2$ globules, forming opal matrix, nm | Wavelength of the maximum intensity of the reflectance spectrum, nm | SLFRS frequency shift, *GHz* | scattering geometry |
|---|---|---|---|---|
| D315 | 315 | 696.1 | 5.1 | forward* |
| D290 | 290 | 645.7 | 6.6 | |
| D270 | 270 | 601.2 | 7.8 | |
| D245 | 245 | 545.5 | 11.1 | |
| D200 | 200 | 445.3 | 13.2 | backward |
| opal matrix filled with acetone | | | | |
| | 200 | - | 12 | backward |
| | 200 | - | 12 | forward* |
| opal matrix filled with ethanol | | | | |
| | 200 | - | 11.7 | backward |
| | 200 | - | 11.1 | forward* |
| | 200 | - | 12 | forward*, T=77 .4 K |

*normal laser beam irradiation

## 3. Experiment

The source of excitation was ruby laser with wavelength of λ=694.3 nm; the duration of the excitation laser pulses was τ=20 ns, spectral width Δν=0.015 cm$^{-1}$.

Plane polarized radiation was emitted with the maximum of energy $E_{max}$ = 0.3 J in almost single axial mode (divergence 3.5 x 10$^{-4}$ rad).

The constant temperature -10°C of ruby rod and the constant power near threshold ensured good repeatability of single laser pulse. The spectral and intensity characteristics of the scattered light beam were measured during each experiment.

The surface of the samples (crystallographic planes family (110) in diamond and growth face of synthetic opal (111)) was irradiated angularly using 3-10 cm focal length lens. Variation of the laser beam focusing allowed to expand the range of the laser intensity for low frequency Raman scattering excitation in the stimulated regime. SLFRS Spectra were measured in 30-60° scattering geometry.

Fabry–Pérot interferometer (FPI) with the range of dispersion 0.7143 cm$^{-1}$ was used for fine spectra resolution.

The experiments were carried out under sub-laser-damage threshold conditions.

Experimental setup for SLFRS excitation is illustrated in Fig. 1.

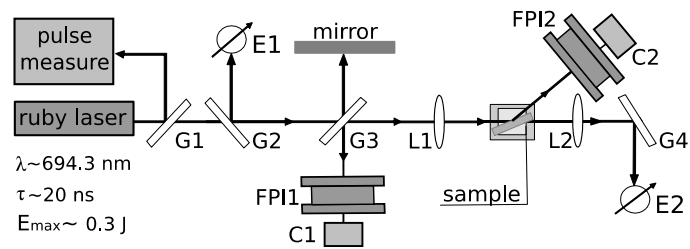

Fig.1. Experimental setup: Ruby laser; G1, G2, G3, G4 –glass plates; E1, E2 – calibrated photodiodes for scattered light energy measurements; mirror– reflecting mirror; FPI1 , FPI2– Fabry–Pérot interferometers for spectral measurements; C1, C2 – CCD cameras; L1, L2 – lenses,  sample – SCD film with a graphitized layer built-in or synthetic opal matrix, pulse measure –laser light parameters measurement system.

The fine structure of the scattered light at 30°-60° and that of the light propagating towards the pump beam (***k***=180° geometry) were measured simultaneously.

SLFRS spectra were registered for SCD films at 30°-60° scattering geometry; no additional spectral components were registered in the forward direction under the experimental conditions.

Additional spectral components appeared in the scattering spectra, when the pump threshold of 0.01 GW/cm$^2$ was exceeded, and it was only for the light reflected at the mirror angle.

The Stokes and anti-Stokes lines for SCD films are defined by the phonon frequency of 1332 cm$^{-1}$ in diamond and appears in our experiments at the wavelengths $\lambda_S$=765 nm and $\lambda_{aS}$=635 nm, respectively in the stimulated Raman scattering spectra (measured by a minispectrometer with a fiber-optic input). Frequency shift of the scattering light from exciting laser light line lied in blue region and had value from 0.28 to 0.30 см$^{-1}$ or 8.4 - 9.3 GHz for all SCD films studied in the experiment (see Table 1). The value of the observed low frequencies differed for different samples and defined by morphology of the samples, especially by the graphitized layer thickness and the restored SCD layer depth of the SCD films (see Table 1).

Figure 2 shows SLFRS spectrum of the light scattered on the submicron single crystal diamond film implanted with 50 keV $^4$He$^+$.

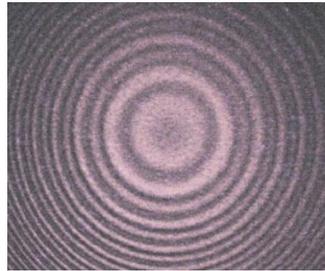

Fig.2. Fabry–Pérot spectrum (dispersion range is 0.7143 cm$^{-1}$) of the radiation scattered on the submicron single crystal diamond film with a buried graphitized layer build in. The depth of the restored SCD layer is 185 nm and the graphitized layer thickness of the sample is 84 nm.

In the case of synthetic opal matrices, additional fine spectral components were observable in majority of the studied samples in the forward direction (see Table 1) for the samples of a few mm thicknesses. Filling of octahedral and tetrahedral pores of the samples with organic liquids made synthetic opals visibly transparent and allowed of the exciting laser light to propagate in the forward direction even for bulk samples.

The value of the SLFRS frequency shift in synthetic opal matrices is of the order of several cm$^{-1}$ (from 5.1 - 13.2 GHz), which corresponds to the vibrations of the GHz range (see Table 2).

Maximum conversion efficiency of the laser light into SLFRS for both type of the samples achieved 40 percent.

For all the investigated samples (both opal matrixes and submicron single crystal diamond films) increasing the average size of the globules in opal matrixes and the thickness of the diamond film lead to decreasing the frequency shift of the SLFRS. It is illustrated in Fig.3.

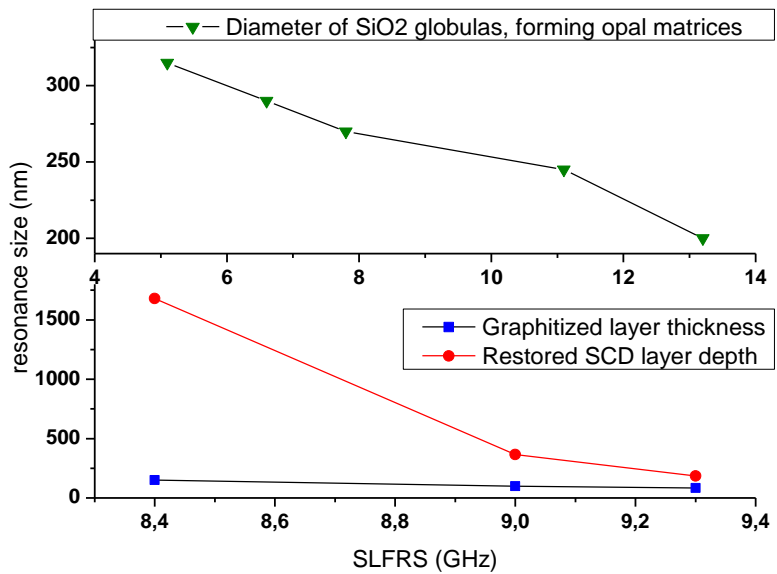

Fig. 3 Morphological dependence of SLFRS frequencies in the samples of synthetic opal matrices, formed by $SiO_2$-globules of a different diameter (upper graph) and in the samples of submicron SCD films with a buried graphitized layer build-in (down graph).

High conversion efficiency and stimulated character of the observed phenomena clearly indicate the effective phonon mode excitation in the studied nanostructured materials –synthetic opal matrices and single crystal diamond films with a buried graphitized layer buil-in.

Resonance phonon mode frequency is defined by the morphology of the samples: the value of the SLFRS spectral shift in SCDs decreases with the thickness of the structure (restored SCD layer depth and graphitized layer thickness) increase (see Table 1). An analogous dependence was observed for samples of synthetic opal matrices. The SLFRS spectral shift in synthetic opals decreases with the diameter of $SiO_2$ globules, forming the opal matrix increase (see Table 2).

Stimulated low frequency Raman scattering, which was observed in gigahertz range in the present experiment is the result of coherent mode excitation due to nonlinear interaction of high-power nanosecond laser pulses with nanosized superstructure of the studied materials.

More specifically, in case of the SCD films, the soft structure of the buried graphitized layer ($sp^2$- carbon) contrasted in the elastic properties and sound velocity with surrounding single crystal diamond lattice. This contrast leads to the absorption of laser excitation by nanosized graphitized build-in layer with further reemitting at the shifted SLFRS frequency.

A similar mechanism works in synthetic opal matrices, where powerful laser pulses amplify eigen vibrations of closely packed $SiO_2$ globules, forming the opal, while electromagnetic laser field induces an electrical dipole moment of the ensemble of the $SiO_2$ globules. This induced polarization is the source of the inelastically scattered wave. Exceeding the threshold value of the laser light leads to the amplification of the initially spontaneously scattered wave and, hence, to an effective growth of the total scattered light.

In both cases nanosized scatters (buried graphitized layer in SCD and monodisperse $SiO_2$ globules, forming synthetic opal matrix) work as acoustic oscillators, which are excited in the media by a powerful laser beam. This leads to the modulation of the pumping light frequency and reemitting of electromagnetic field by nanoparticles (nanolayer) vibrations at red-shifted frequency.

## 5. Summary

Our experiments demonstrate universal nature of stimulated low frequency Raman scattering for different type of structures. SLFRS shows the morphological dependence for synthetic opal matrices, where SLFRS frequency shifts lie in GHz region and have negative relationship with the period of nanosized superstructure, as for as submicron SCD films with built-in graphitized layer, where gigahertz SLFRS frequencies are also in negative dependence with geometry of nanosized films.

Submicron single crystal diamond films with graphitized built-in layer of nanosized thickness and bulk synthetic opals formed by monodisperse dielectric globules are proposed to be used in acoustooptical devices for spectroscopic and environmental applications as scalable reproductive Raman active mediums, matching criteria of compatibility .

**Acknowledgments**

This work was supported by the Russian Foundation for Basic Research (projects no. 16-32-60026 and no. 15-02-02875).